\documentclass[prl,twocolumn, amsmath, superscriptaddress,floatfix]{revtex4}
\usepackage{graphicx}
\usepackage{bm}

\newcommand{\ket}[1]{\ensuremath{|{#1}\rangle}}

\newcommand{\ts}[2]{\ensuremath{{#1}_{\text{\lowercase{#2}}}}}
\newcommand{\tsc}[2]{\ensuremath{{#1}_{\textsc{\lowercase{#2}}}}}
\renewcommand{\vec}[1]{\mathbf{#1}}  

\newcommand{\Ginzton}{
    \affiliation{Edward L. Ginzton Laboratory, Stanford University,
    Stanford, California 94305-4088, USA}}
\newcommand{\NII}{\affiliation{National Institute of Informatics, 2-1-2 Hitotsubashi,
Chiyoda-ku, Tokyo 101-8430, Japan}}

\begin{document}
\title{Ultrafast optical spin echo for electron spins in semiconductors}

\author{Susan M. Clark}\Ginzton
\email[Electronic address: ]{sclark4@stanford.edu}
\author{Kai-Mei C. Fu}
\affiliation{Quantum Science Research, Hewlett-Packard
Laboratories, 1501 Page Mill Road, MS1123, Palo Alto, California
94304}
\author{Qiang Zhang}\Ginzton\NII
\author{Thaddeus D. Ladd}\Ginzton\NII
\author{Colin Stanley}
\affiliation{Department of Electronics and Electrical Engineering, Oakfield Avenue, University of Glasgow, Glasgow, G12 8LT, United Kingdom}
\author{Yoshihisa Yamamoto}\Ginzton\NII

\begin{abstract}
Spin-based quantum computing and magnetic resonance techniques
rely on the ability to measure the coherence time, $T_2$, of a
spin system.  We report on the experimental implementation of
all-optical spin echo to determine the $T_2$ time of a
semiconductor electron-spin system.  We use three ultrafast
optical pulses to rotate spins an arbitrary angle and measure
an echo signal as the time between pulses is lengthened. Unlike
previous spin-echo techniques using microwaves, ultrafast
optical pulses allow clean $T_2$ measurements of systems with
dephasing times ($T_2^*$) fast in comparison to the timescale
for microwave control.  This demonstration provides a step
toward ultrafast optical dynamic decoupling of spin-based
qubits.
\end{abstract}

\maketitle

Proposals for spin-based quantum information processors have
generated renewed interest in the coherent control and
decoherence of electron spins in many environments. Determining
the decoherence time, $T_2$, for spins has been particularly
important in semiconductors since it sets the timescale at
which error correction must occur in a semiconductor-spin-based
quantum computer~\cite{Loss1998,Imamoglu1999,Clark2007}.  More
generally, the measurement of $T_2$ provides information about
noise processes in a spin's environment, which forms the basis
for many techniques in magnetic resonance spectroscopy and
magnetic resonance imaging.

In some systems, however, measurement of the $T_2$ decoherence
time is obscured by inhomogeneities in spin environments.  Spin
inhomogeneity causes a perceived loss of coherence, or static
dephasing, on a much faster timescale called $T_2^*$.  For an
ensemble of spins, this inhomogeneity is due to each spin
having a different Larmor frequency due its local
environment~\cite{Fu2005, Dutt2006}.  For a single spin, this
inhomogeneity is caused by slow environmental changes during
temporal averaging~\cite{Petta2005, Koppens2008, Hanson2008}.
Techniques exist, however, to decouple spins from these
inhomogenous environments~\cite{Slichter1996}, but they rely on
the coherent manipulation of the spin.  If $T_2^*$ is fast
compared to the time required to perform this manipulation,
these techniques become ineffective.  One way around this
problem is to use a faster method of spin manipulation.  Here,
we use ultrafast optical pulses~\cite{Gupta2001, Economou2006,
Dutt2006, Fu2008, Berezovsky2008, Press2008} to perform the
spin manipulations instead of slower microwave pulses.
Ultrafast optical pulses not only allow the efficient
refocusing of static dephasing caused by inhomogeneities, but
may also provide sufficient speed to enable the dynamic
decoupling of spins from the noise sources that cause $T_2$
decoherence~\cite{Viola1998, Witzel2007, Yao2007, Uhrig2007,
Lee2008}.

\begin{figure}[hb!]
\includegraphics[width = \columnwidth]{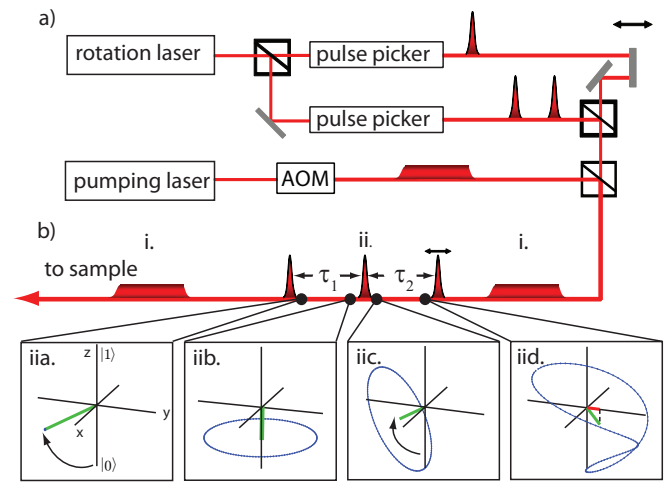}
\caption{(Color online) {\bfseries a)} Experimental setup to generate necessary pulses for all optical spin echo.  {\bfseries b)} Pulse sequence for optical spin echo experiment.  The labels i and ii, match those in Fig.~\ref{energylevels}b.  Panels below pulse sequence show the orientation of an ensemble of inhomogenous spins (blue dots) on the Bloch sphere at different points during a $\pi/3 - \pi/3 - \pi/3$ pulse sequence.  The green line shows the average spin polarization vector and the thick red line in panel 2d represents the $y$-axis projection of the average spin vector.  After the final ultrafast pulse and pumping pulse, this projection will be the spin polarization that we observe. }\label{pulsesequence}
\end{figure}

The simplest method to measure $T_2$ is the Hahn spin-echo
sequence~\cite{Hahn1950}, which consists of a $\pi/2$ pulse, a
period of free evolution, a $\pi$ pulse, and then another,
equal period of free evolution.  This sequence is a workhorse
technique in the fields of nuclear magnetic resonance (NMR) and
electron spin resonance (ESR).  Recently, this technique has
been used to measure $T_2$ for potential solid-state spin
qubits~\cite{Jelezko2004, Petta2005, Hanson2006, Hanson2008,
Koppens2008}. Another, more indirect way to measure $T_2$ is
via pulsed mode-locking.  This technique may be applied in some
short-$T_2^*$ systems using trains of ultrafast optical pulses
and a $T_2$ may be extracted by comparing the amount of
generated coherence for different pulse train repetition
times~\cite{Greilich2006}.  This mode-locking technique has 
recently been modified to observe optical spin echo
signals~\cite{Greilich2009}, however, no measurement of $T_2$
has yet been made with these echo signals.  Additionally, the
coherences preserved by modelocking are determined by the
polarization of the control pulses, complicating the extension
of this technique to the preservation of arbitrary, possibly
entangled, spin qubit states.

The Hahn spin-echo technique may be generalized to arbitrary
spin-rotation angles.  A spin rotation of angle $\theta$ due to
an ultrafast pulse is modeled as an instantaneous appearance of
a large effective field along the $x$-axis that is much larger
than the applied magnetic field in the $z$-direction.  As a
result, the rotation is described by an instantaneous unitary
rotation operator $R_x(\theta) = \exp(-i\theta S^x)$ for the
Pauli spin-operator $S^x$ ($S = 1/2)$. Similarly, free rotation
about the magnetic field is modeled by a unitary rotation
operator about the $z$-axis, $R_z(\theta)$, where the angle of
rotation is determined by $\theta = \omega_n \tau$, where
$\tau$ is the time of free procession and $\omega_n =
g\tsc\mu{B} B^z(\textbf{r}_n)/\hbar$ is the Larmor frequency of
the $n^{\text{th}}$ spin.  Here, $g$ is the gyromagnetic ratio for
the material, $\tsc\mu{B}$ is the Bohr magneton, and
$B^z(\vec{r}_n)$ is the magnetic field at each spin position
$\vec{r}_n$.  The Larmor frequency varies from spin to spin
primarily due to random nuclear hyperfine fields.  Assuming we
begin with the $n^{\text{th}}$ spin initialized in the spin
down state, $|\downarrow\rangle$, the final state of the system, $\ket{\ts\psi{f}}_n$, after
the sequence of three ultrafast rotations of angles
$\theta_1$,$\theta_2$, and $\theta_3$, separated by two
intervals of free precession for times $\tau_1$ and $\tau_2$, can be described by
\begin{equation}
\ket{\ts\psi{f}}_n  = R_x(\theta_3)R_z(\omega_n \tau_2)R_x(\theta_2)R_z(\omega_n \tau_1)R_x(\theta_1)\ket{\downarrow}.
\end{equation}
In our experiment, we measure probability $P$ that a spin is
flipped to the spin up state $\ket{\uparrow}$ averaged over $N$
spins.  To model inhomogeneity, we suppose that the Larmor
frequencies have a Gaussian probability distribution, so this
average projection becomes
\begin{eqnarray}
P&=&\frac{1}{N}\sum_{n=1}^N |\langle\uparrow\!|\ts\psi{f}\rangle_n|^2\nonumber\\
&=&\frac{1}{\sqrt{2\pi}\sigma}\int_{-\infty}^\infty d\omega
|a(\omega)|^2e^{-(\omega-\omega_0)^2/2\sigma^2},
\end{eqnarray}
where $\omega_0$ is the mean and $\sigma\sim 1/T_2^*$ is the
width of the spin resonance distribution.  The average
projection onto the $z$-axis is then found to be
\begin{align}\label{bigz}
\langle \sigma^z&(\tau_1,\tau_2) \rangle = 2P-1
 =\\
 &-\cos\theta_3\cos\theta_2\cos\theta_1\label{aterm}
\notag\\
 &+\cos\theta_3\sin\theta_2\sin\theta_1\cos\omega_0\tau_1e^{-\frac{1}{2}\sigma^2\tau_1^2}
\notag\\
 &+\sin\theta_3\sin\theta_2\cos\theta_1\cos\omega_0\tau_2e^{-\frac{1}{2}\sigma^2\tau_2^2}
\notag\\
 &+ \sin\theta_3\cos^2\frac{\theta_2}{2}\sin\theta_1\cos[\omega_0(\tau_1+\tau_2)]e^{-\frac{1}{2}\sigma^2(\tau_1+\tau_2)^2}
\notag\\
 &-
 \sin\theta_3\sin^2\frac{\theta_2}{2}\sin\theta_1\cos[\omega_0(\tau_1-\tau_2)]e^{-\frac{1}{2}\sigma^2(\tau_1-\tau_2)^2}.
\notag\end{align}

Most of the terms in Eq.~\ref{bigz} result in polarizations
that rapidly vanish due to $T_2^*$ dephasing as the time
between pulses is increased, but the last term describes a
spin-echo which is immune to static $T_2^*$ dephasing.  This
term is maximized for the Hahn echo condition $\theta_1 =
\theta_3 = \pi/2$ and $\theta_2 = \pi$.  The Hahn echo
condition, however, is not required to see the echo; there will
still be a finite polarization for almost any angle of the
three pulses.  For one visualization of how the spins refocus
in the small angle case, refer to the panels in
Fig.~\ref{pulsesequence}. This generalization is particularly
important when using ultrafast pulses, because the decoherence
induced by the pulses themselves may increase with
angle~\cite{Fu2008}.

\begin{figure}
\includegraphics[width = \columnwidth]{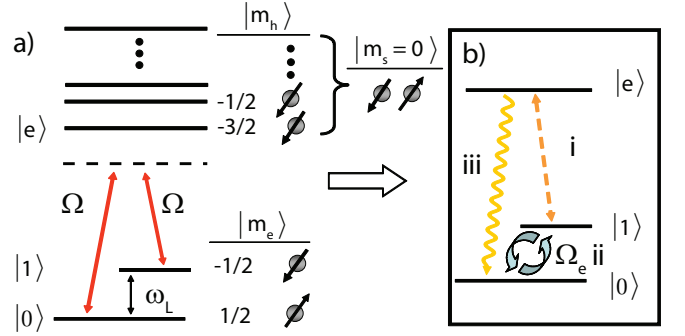}
\caption{(Color online) Energy level diagram for a donor bound exciton system in a magnetic field.
{\bfseries a)} Ground states $|0\rangle$ and $|1\rangle$ are split in energy $\hbar \omega_L$
by a magnetic field.  They are optically connected to the donor bound exciton states, $|\text{e}\rangle$ via a detuned, ultrafast pulse of Rabi frequency $\Omega$.  Spins of the relevant levels are shown to the right of the diagram.  {\bfseries b)} Relevant energy levels and applied optical fields in the experiment.  The pumping laser applies electric fields on resonance to pump population into the $|0\rangle$ state (orange dashed line, i).  The rotation laser performs rotations between the ground states with effective Rabi frequency $\ts\Omega{e}$ (blue arrows, ii).  The pumping laser applies another field on resonance (i) and light is collected from the $|0\rangle \leftrightarrow |\text{e}\rangle$ transition (yellow line, iii)}\label{energylevels}
\end{figure}
We applied this sequence of three optical rotations to an
ensemble of electron spins bound to neutral Si donors in
GaAs~\cite{Karasyuk1994}.  For this material, high pulse powers
induce additional decoherence, limiting $\theta$ to values less
than about $\pi/3$~\cite{Fu2008}.  An energy level diagram of
this system in an applied magnetic field is shown in
Fig.~\ref{energylevels}a.  The ground states are formed by a
single electron spin bound to a Si donor in GaAs. They are
denoted $|0\rangle$ and $|1\rangle$ depending on the spin of
the electron and are split by an energy $\hbar\tsc\omega{l}
\approx
50$~GHz via a magnetic field, $B_{\text{ext}} = 10$~T.
These ground states are optically connected to the
donor-bound-exciton (D$^0$X) state (denoted
$|\text{e}\rangle$), which consists of an additional
electron-hole pair.  Despite the many excited states of the
donor-bound-exciton system, we can approximate the ground
states as a two-level system via adiabatic elimination of the
excited states (valid if the detuning, $\Delta = 1$~THz, is
much larger than other rates in the system).  The ultrafast
pulse (2 ps) couples states $|0\rangle$ and $|1\rangle$ with an
effective Rabi frequency, $\ts\Omega{e}\sim\Omega^2/\Delta$,
where $\Omega$ is the instantaneous optical Rabi
frequency~\cite{Clark2007, Fu2008, Press2008}.  For our
experiments, the pulse width of the ultrafast pulse is held
constant, so the power of the pulse
determines the spin rotation angle.

The experimental setup to generate this pulse sequence is shown
in Fig.~\ref{pulsesequence}a and the pulse sequence used to
perform the spin echo measurement is depicted in
Fig.~\ref{pulsesequence}b.  First we initialize the spins to
the $|0\rangle$ state by applying a long pulse from a
continuous-wave ring laser (labeled ``pumping laser" in
Fig.~\ref{energylevels}a) gated by an acousto-optic modulator
(AOM) resonant on the $|1\rangle \leftrightarrow |\text{e}\rangle$
transition (labeled i in Fig.~\ref{pulsesequence}b and Fig.~\ref{energylevels}b).  The first and second rotation
pulses are picked from a pulse train of a Ti:Sapphire
modelocked laser (``rotation laser") by a pulse picker
(electro-optic modulator (EOM) followed by an AOM for improved
extinction ratio) and are separated by $\tau_1$, which is
always a multiple of the repetition time of the laser
(13.2~ns).  The third pulse arrives at a time $\tau_2$ after
the second pulse, which is determined by both a pulse picker
and an optical delay line which can vary $\tau_2$ by 10s of picoseconds.  In order
to measure rephasing, the difference between the free evolution
times, $|\tau_2 - \tau_1|$, is kept smaller than the 1~ns
$T_2^*$ dephasing time~\cite{Fu2005}.  We determine the final
state of the system by projecting the rephased spin
polarization vector onto the $z$-axis with the application of
another long pulse ($\mu$s) from the pumping laser on resonance
with the $|1\rangle \leftrightarrow |\text{e}\rangle$ transition
(labeled i in Fig.~\ref{pulsesequence}b and
Fig.~\ref{energylevels}b).  We monitor the spontaneous
emission at the frequency of the $|0\rangle \leftrightarrow
|\text{e}\rangle$ transition (labeled iii in
Fig.~\ref{energylevels}b), which will be proportional to the
amount of population in state $|1\rangle$ .  We filter out pump
laser scatter from the spontaneous emission signal via a
monochromator and polarization selection rules.  The final
signal is measured with an avalanche photodiode.

For these ultrafast rotations, only the arrival time of the
pulse determines the phase of the rotation~\cite{Clark2007}, so
the optical phase of the pulse does not need to be externally
stabilized.  However, the pulse arrival time needs to be
stabilized to much less than the Larmor frequency of the spin
system.  The jitter of the modelocked Ti:Sapphire laser was
measured using a technique described by von der
Linde~\cite{vonderLinde1986} and was found to be about 1~ps
over a millisecond timescale.  By keeping $\tau_1$ and $\tau_2$
less than a millisecond, individual pulses picked from the
modelocked train can be used for this pulse sequence without
external stabilization.

\begin{figure}[t]
\includegraphics[width = \columnwidth]{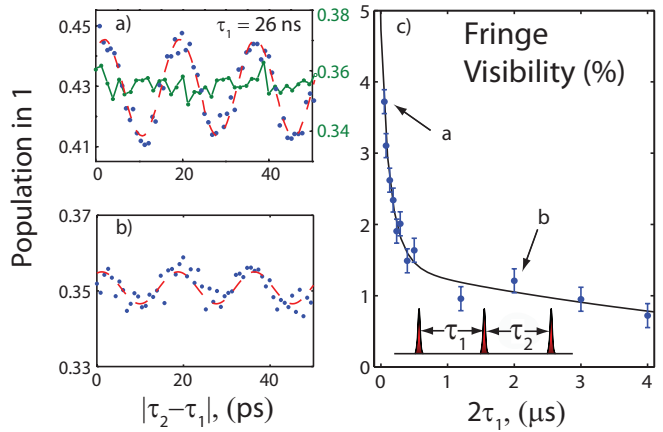}
\caption{(Color online) Results of all-optical spin echo experiment {\bfseries a)} Population in $|1\rangle$ as second pulse is scanned in a two pulse experiment where $\tau_1 = 26$~ns (solid green line).  No fringes can be seen, indicating complete dephasing.  The blue dots and dotted red curve are data and a fit respectively to a three pulse experiment with $\tau_1 = 26$~ns.  For the three pulse experiment, there are clear oscillations.  {\bfseries b)} Population in $|1\rangle$ as $\tau_2$ is scanned for $2\tau_1 = 2~\mu$s (blue dots) and the fit to a sine function (red dashed line).  The visibility is 1.2~\%. {\bfseries c)} Fringe visibility vs. pulse separation between first and last pulse, $2\tau_1$.  Error bars are calculated from the standard deviation of several measurements of the fringe visibility at one point.  The fit is a phenomenological decay (black solid line) that includes a pulse induced decoherence and an intrinsic $T_2$ decoherence with a decay time of 6.7~$\pm~2.5~\mu$s.   The points shown in a) and b) are labeled.}\label{T2data}
\end{figure}
The results of the experiment can be seen in Fig.~\ref{T2data}.
When we apply only two pulses, as in a Ramsey fringe
experiment, we see no fringes (Fig.~\ref{T2data}a, solid green line)
for pulse separations greater than 26 ns, indicating that the
spins are completely dephased.  When we add a third pulse, we
recover fringes that have a frequency equal to the Larmor
frequency (blue dots and red dashed line in Fig.~\ref{T2data}a
and b) which is evidence of the echo.  Some fluctuations in the
experiment cause a small linear drift of random direction and
amplitude in the overall detector count-rate.  We remove the
drift by fitting each fringe curve to a sum of terms constant,
linear, and oscillatory in $\tau_2$, and then subtracting the
linear term.  We calculate the visibility for each fringe curve
as $V = (\text{max}-\text{min})/(\text{max}+\text{min})$, where
$\text{max}$ and $\text{min}$ are the maximum and minimum of
the fitted sine curve.  In Fig.~\ref{T2data}c we see the decay
of the fringe visibility with pulse separation, indicating
decoherence in this system.

The data in Fig.~\ref{T2data}c are well described by a phenomenological model with two sources of decoherence.  The first source is intrinsic to the sample and leads to a decay of the echo with rate $T_2^{-1}$.  The second source is generated by the pulses themselves. It is known that the rotation pulses have a finite fidelity due to a decoherence source generated by the pulses. It was found in Ref.~\cite{Fu2008} that as more pulse power is applied, there appears a new dephasing of the bound-exciton state that is proportional to pump power. This dephasing may be due to local heating resulting from background absorption of the laser, which induces some population of phonons or other excitations, such as excited impurity states.  In our model, these pulse-induced remnant excitations also decohere the spins at a rate proportional to their population. Since the spurious excitation is assumed to equilibrate at some timescale $\ts{T}{h}$, we presume this effect vanishes in time as $\exp(-t/\ts{T}{h})$, and so a coherence $\langle \sigma^+ \rangle$ obeys a Bloch equation of the form
\begin{equation}
\frac{d}{dt}\langle \sigma^+(t) \rangle_n = \left[i\omega_n-\frac{1}{T_2}-Re^{-t/\ts{T}{h}}\right]\langle \sigma^+(t) \rangle_n,
\end{equation}
which may be analytically integrated.

Using this decoherence model, the visibility for $\tau_1=\tau_2=\tau \gg T_2^*\sim 1/\sigma$ is
\begin{equation}
V(\tau) = V_0 e^{-2\tau/T_2-2R\ts{T}{h}[1-\exp(-\tau/\ts{T}{h})]}.
\end{equation}
This curve is fit to the visibility data in Fig.~\ref{T2data}c (black line).  The resulting fitting parameters are the intrinsic decoherence time $T_2 = 6.7 \pm 2.5~\mu\text{s}$; the pulse-induced decoherence time $R^{-1}= 175 \pm 30$~ns; the relaxation rate of the excitations contributing to pulse-induced decoherence $\ts{T}{h}=100\pm 20$~ns; and the initial visibility before decay $V_0=0.047 \pm 0.003.$ This last parameter $V_0$ may be estimated from Eq.~(\ref{bigz}) as
\begin{equation}
V_0=
    \langle \sigma^z \rangle_0 \frac{D(\theta_1)D(\theta_2)\sin(\theta_3)\sin(\theta_2^2/2)\sin(\theta_1)}
    {1-\langle\sigma^z\rangle_0\cos(\theta_3)\cos(\theta_2)\cos(\theta_1)},
\end{equation}
where $\langle \sigma^z\rangle_0$ is the initial polarization created by optical pumping and $D(\theta)$ is the amount of spin decoherence created by a single pulse of angle $\theta$.  Both of these may be estimated from previous studies~\cite{Fu2008} as $D(\theta)\approx 1-0.25\theta$ and $\langle \sigma^z \rangle\approx 0.9$, leading to a $V_0$ of 5~\%, in agreement with the fit.

The measured $T_2 = 6.7 \pm 2.5~\mu\text{s}$ is consistent with
previous measures of $T_2$ in quantum dot systems using
microwave spin echo~\cite{Petta2005, Koppens2008} and
mode-locking~\cite{Greilich2006}.  A microsecond $T_2$,
however, is still orders of magnitude shorter than the
theoretical maximum of twice the spin relaxation time
(2$T_1$)~\cite{Coish2004}, which has been measured to be
milliseconds in this sample~\cite{Fu2006}.  This observation
indicates that there are sources of dynamic decoherence, most
likely nuclear spin diffusion, which has been theoretically
predicted to cause coherence decay on this
timescale~\cite{Paget1982, deSousa2003b, Yao2006}.  This type
and other types of dynamic decoherence, however, can be
reversed using a series of $\pi$-pulses applied faster than the
characteristic timescale of
decoherence~\cite{Viola1998,Witzel2007, Yao2007, Uhrig2007,
Lee2008}, therefore extending the $T_2$ time of the system.  In
cases where $\pi$-pulses are not possible, a series of
small-angle pulses can be applied over several Larmor periods
to sum to a $\pi$-pulse~\cite{Fu2008}.  Because optical pulses
can be applied more quickly than microwave pulses, the
demonstration in this paper is a critical step towards
eliminating decoherence that occurs on a fast timescale,
potentially making many more materials suitable for
applications requiring long $T_2$.

In conclusion, we have demonstrated that three ultrafast
optical pulses can produce an echo signal to measure the $T_2$
time of electron-spin systems.  Although here we applied this
technique to Si:GaAs, we believe it has potential applications
in other materials relevant to magnetic resonance, particularly
those with fast decoherence times.  We have also demonstrated
that partial rephasing is possible even when $\pi$ and $\pi/2$
pulses are unavailable, meaning the technique can be applied
when a system has a small optical dipole moment, when only low
laser power is available, or when optical dephasing is present.
Lastly, optical pulses have the potential to extend the
decoherence time in semiconductor systems by dynamical
decoupling, allowing more materials to be suitable for
application that require long $T_2$ times. Such a scheme could
be used to extend the spin-memory time of a spin-based quantum
computer and can be integrated into quantum bus schemes for
quantum computing~\cite{Clark2007}.

S. M. C. was partially supported by the HP Fellowship Program
through the Center for Integrated Systems.  This work was
financially supported by the MURI Center for photonic quantum
information systems (ARO/ARDA Program DAAD 19-03-1-0199),
JST/SORST program for the research of quantum information
systems for which light is used, University of Tokyo Special
Coordination Funds for Promoting Science and Technology, and
MEXT, NICT.

\end{document}